\documentclass[twocolumn, 
prb,preprintnumbers,amsmath,amssymb,floatfix]{revtex4}

\usepackage[linktocpage,bookmarksopen,bookmarksnumbered]{hyperref}
\usepackage{graphicx}
\usepackage{dcolumn}

\usepackage{amsmath,graphics,epsfig,color,verbatim,ulem}

 \newcommand{\vk}{{\mathbf{k}}}

\begin{document}

\setlength{\pdfpageheight}{\paperheight}
\setlength{\pdfpagewidth}{\paperwidth}

\title{Magnetism and Charge Dynamics in Iron Pnictides}
\author{Z. P. Yin}
\author{K. Haule}
\author{G. Kotliar}
\affiliation{Department of Physics and Astronomy, Rutgers University, Piscataway, NJ 08854, United States.}
\date{\today}
\maketitle

\textbf{
In a wide variety of materials, such as layered copper oxides, 
heavy fermions, organic salts, 
and the recently discovered iron pnictides, 
superconductivity is found in close proximity to a magnetically ordered state\cite{Intro1,Intro2}. 
The character of the proximate magnetic phase is thus believed 
to be crucial for understanding the differences between 
the various families of unconventional superconductors and the mechanism of superconductivity.
Unlike the antiferromagnetic order in cuprates, 
which is well described by the spin Heisenberg model, 
the nature of the magnetism and of the underlying electronic state 
in the iron pnictide superconductors is not well understood.
Neither density functional theory 
nor models based on atomic physics and superexchange, 
account for the small size of the magnetic moment~\cite{Ba-m}.
Many low energy probes such as transport~\cite{Chu}, 
scanning tunneling microscopy~\cite{Chuang} 
and angle-resolved photoemission spectroscopy (ARPES)[\onlinecite{Richard}] 
measured strong anisotropy of the electronic states 
akin to the nematic order in a liquid crystal, 
but there is no consensus on its physical origin, 
and a three dimensional picture of electronic states 
and its relations to the optical conductivity in the magnetic state is lacking.
Using a first principles approach, 
we obtained the experimentally observed magnetic moment, 
optical conductivity, 
and the anisotropy of the electronic states.
The theory connects ARPES, 
which measures one particle electronic states, 
optical spectroscopy, 
probing the particle hole excitations of the solid and neutron scattering 
which measures the magnetic moment.
We predict a manifestation of the anisotropy in the optical conductivity, 
and we show that the magnetic phase arises from the paramagnetic phase 
by a large gain of the Hund's rule coupling energy and a smaller loss of kinetic energy, 
indicating that iron pnictides represent a new class of compounds where the nature of magnetism 
is intermediate between the spin density wave of almost independent particles, 
and the antiferromagnetic state of local moments.
}

Below the Neel temperature of the order of $150\,$K the parent
compounds of the iron pnictide superconductors remain metallic with a
magnetization density oscillating in space (spin density wave, SDW).
The sublattice magnetization is concentrated on iron atoms and 
its arrangement in space is antiferromagnetic in the $x$ direction and
ferromagnetic in the $y$ direction\cite{Ba-lattice}.

We use the combination of the density functional theory (DFT) 
with local density approximation (LDA) and the
dynamical mean field theory (DMFT)[\onlinecite{review}] to study the
archetypical iron pnictide compound BaFe$_2$As$_2$ in both the
magnetic SDW and the paramagnetic (PM) state.
The size of the theoretical magnetic moment is 0.86~$\mu_B$, similar
to the measured moment of 0.87 $\mu_B$[\onlinecite{Ba-m}], but much
smaller than 1.75~$\mu_B$[\onlinecite{Yin-antiphase}] obtained 
by the local spin density approximation (LSDA) within DFT. 
The considerably smaller magnetic moment obtained here is due to the
fact that the competing PM metallic state is a correlated
metal, which contains very fast fluctuating moments in time, but no
static moment. Only a small part of these fluctuating moments acquires
a static component in the ordered state.

\begin{figure}[bht]
\includegraphics[width=0.75\linewidth]{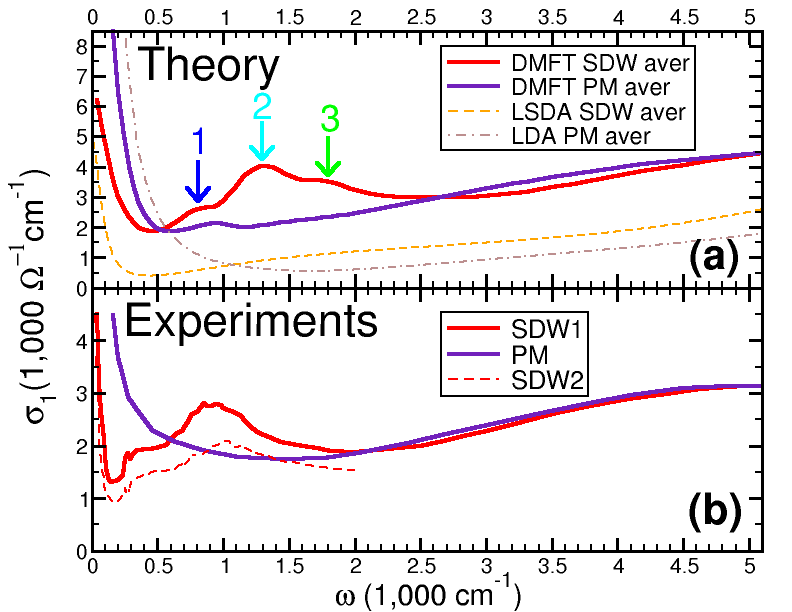}
\includegraphics[width=0.75\linewidth]{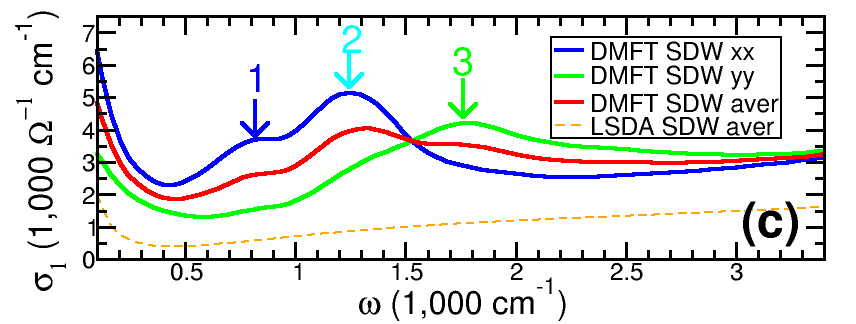}
\includegraphics[width=0.75\linewidth]{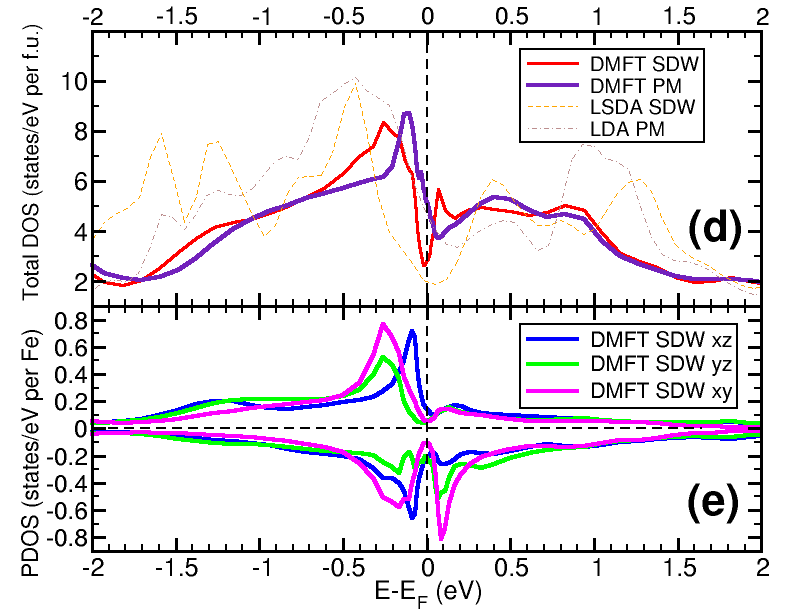}
\caption{
Optical conductivity and DOS of BaFe$_2$As$_2$. (a)calculated in-plane average optical conductivity
by LDA+DMFT and L(S)DA in the SDW and PM states; (b)experimental in-plane optical conductivity
in the SDW and PM states. Data are taken from Refs.[\onlinecite{Hu}] and [\onlinecite{Nakajima}];
(c)the $xx$, $yy$, and in-plane average optical conductivity in the SDW state calculated by LDA+DMFT and LSDA (only the average is shown for LSDA);
(d)total DOS in the SDW and PM states calculated by LDA+DMFT and L(S)DA
and (e)the projected DOS of Fe $3d$ $xz$, $yz$ and $xy$ orbitals in the SDW state calculated by LDA+DMFT,
plotting positive/negative for majority/minority spin.
}
\label{optical-conductivity}
\end{figure}

The onset of magnetic order has a profound impact on the electronic
structure, and these changes are probed by optical spectroscopy.
Figure~\ref{optical-conductivity}(a) shows the in-plane (averaged over
$x$ and $y$ direction) optical conductivity of BaFe$_2$As$_2$ in the
SDW and PM states calculated by both LDA+DMFT and
L(S)DA. Fig.~\ref{optical-conductivity}(b) reproduces measured in-plane optical
conductivity from Refs.~\onlinecite{Hu} and \onlinecite{Nakajima}.
Both theory and experiments\cite{Hu,Nakajima} show a reduction of the
low frequency Drude peak, which indicates a removal of a large
fraction of carriers in the ordered state.
Our calculation captures all the important qualitative features
measured in experiments.
In both PM and SDW states there is a broad peak due to interband
transitions centered around $5500\,$cm$^{-1}$[\onlinecite{Kutepov}]. Below
2000$\,$cm$^{-1}$ the optical conductivity of the SDW phase shows a
few extra excitations appearing as peak and shoulders centered at 1250
cm$^{-1}$ (arrow 2 in cyan), shoulder structure at about 800 cm$^{-1}$
(arrow 1 in blue), and a small peak at 1800 cm$^{-1}$ (arrow 3 in
green).
These additional excitations appear in experiment at slightly smaller
energies, as seen in Figs.~\ref{optical-conductivity}(a) and (b).

These extra peaks and shoulders strongly depend on the polarization of
the light, as shown in Fig.~\ref{optical-conductivity}(c), where we
plot separately the $x$ and the $y$ component of the optical
conductivity. The first two excitations (blue arrow 1 and cyan arrow
2) are much more pronounced in the $x$ direction, while the third peak
(green arrow 3) is more pronounced in the $y$ direction. Also the
conductivity is considerably larger in the $x$ direction (antiparallel
spins) than in the $y$ direction (parallel spins).
At low frequency, the optical conductivity of the SDW phase shows
Drude like behavior, with Drude weight considerable smaller than in
the PM phase. The theoretical value of the plasma frequency for the
$x$, $y$ and $z$ direction ($1.14$, $0.88$ and $0.69\,$eV,
respectively) are much smaller than the plasma frequency of $1.60\,$eV
of the PM DMFT calculation~\cite{Kutepov}, which agrees very well with
experimental estimates~\cite{Hu, Chen}.
This reduction was also observed experimentally~\cite{Hu}.
The anisotropy of the in-plane optical conductivity has not been
studied experimentally, because of the fairly coarse spatial
resolution imposed by the diffraction limit, spanning multiple
magnetic domains. Near field optics\cite{Basov}, or techniques to
prepare monodomain samples~\cite{Chu}, are promising avenues to test
our prediction.

It is useful to analyze the optical conductivity at various frequency
scales. 
The Drude weight is controlled by
the Fermi surface size, and by the mass enhancement of the low energy
quasiparticles. In the SDW state, the mass enhancement is
smaller ($2$, $1.7$,
$1.7$ and $1.5$ for the $t2g$/majority, $t2g$/minority, $eg$/majority,
and $eg$/minority orbitals) than in the PM state 
( $3$
and $2$, for the $t2g$ and $eg$ orbitals, respectively).
While the quasiparticles become lighter in the SDW phase, the Fermi
surface area is much smaller, and the latter effect dominates,
resulting in a reduction of the Drude weight.

We also integrated the in-plane optical conductivity to obtain the
effective kinetic energy of a low energy model, in both the SDW phase
and the PM phase. At very low energies, the onset of magnetism results
in an \textit{increase} of the optical conductivity, due to the
coherence-incoherence crossover. The long range order makes the
material more coherent. Consequently the very
low energy model gains kinetic energy in the ordered state.
At intermediate energies, however, 
kinetic energy is lost as the result of the opening of the SDW gap on
the Fermi surface.

Since the optical conductivity of the PM phase is quite
temperature dependent, we compare the SDW phase and PM phase at the
same temperature ($T=72.5$K), the latter being a metastable state at low
temperature. Using this procedure, we find that the missing weight
from opening the SDW gap is recovered only around 10000$\,$cm$^{-1}$,
many times larger scale than the gap value. This should be contrasted
with the classic weakly correlated materials, where the spectral
weight is recovered immediately above the SDW gap.

Recent dc conductivity and scanning tunneling microscopy (STM) measurements have detected a large
anisotropy in the $ab$ plane. 
\cite{Chu} \cite{Chuang}. 
The theoretical zero
frequency limit of the optical conductivity is 7350, 4400, and
3350 $\Omega^{-1} cm^{-1}$ for the $xx$, $yy$ and $zz$ components,
respectively. This implies that the resistivity in the $y$ direction
is about 1.67 times of that in the $x$ direction, in good agreement
with the results of resistivity measurements of Chu {\it et al.}~\cite{Chu}.

We now turn to electronic density of
states. Figure\ref{optical-conductivity}(d) shows the total density of
states (DOS) in both the PM and SDW phases. In the SDW phase, DMFT DOS
shows a clear pseudogap on the scale of 0.15$\,$eV around the Fermi
level, in good agreement with STM measurements~\cite{Chuang}. 
The LSDA DOS also shows a pseudogap at the Fermi level~\cite{Yin-prl},
however, its width is more than 0.5 eV, therefore LSDA misses the
structure below 4000 $cm^{-1}$ in the optical conductivity.

The onset of stripe magnetic phase is also accompanied by a
rearrangement of the iron crystal field states, which gives rise to
orbital polarization. This polarization is uniform in space
(ferro-orbital ordering), as surmised by Singh\cite{RSingh}. The
partial density of states of an Fe atom is shown in
Fig.\ref{optical-conductivity}(e). The minority density is given a
negative sign. To extract the anisotropy of the electronic structure,
we integrate the partial density of states of $xz$ and $yz$ orbital
($A_{xz}(\omega)$ and $A_{yz}(\omega)$) to
obtain their occupation, and evaluate their difference $\Delta
n(\Lambda)=\int_{-\Lambda}^0[A_{xz}(\omega)-A_{yz}(\omega)]d\omega/(\frac{1}{2}
\int_{-\Lambda}^0[A_{xz}(\omega)+A_{yz}(\omega)]d\omega)$. This
defines the energy dependent orbital polarization. For large cutoff
$\Lambda$, the orbital polarization is close to zero for majority
electrons and around 0.13 for minority electrons. At low energy, the
anisotropy is enhanced to $1.23$ ($0.45$) for majority (minority)
carriers, when $\Lambda$ is $0.15\,$eV, the size of the optical SDW gap.

The anisotropy of the partial density of states provides a natural
explanation for the anisotropy in the optical conductivity; the $yz$
density of states has less electronic states at the Fermi level and
the main peak of the $yz$ orbital is further away from the Fermi level
compared to $xz$ orbital. The optical conductivity in $x$ direction
comes primarily from $xz$ and $xy$ orbitals, and is thus larger than
the conductivity in $y$ direction, which is connected to $yz$ and $xy$
orbitals.

\begin{figure}[!ht]
\centering{
  \includegraphics[width=1.0\linewidth]{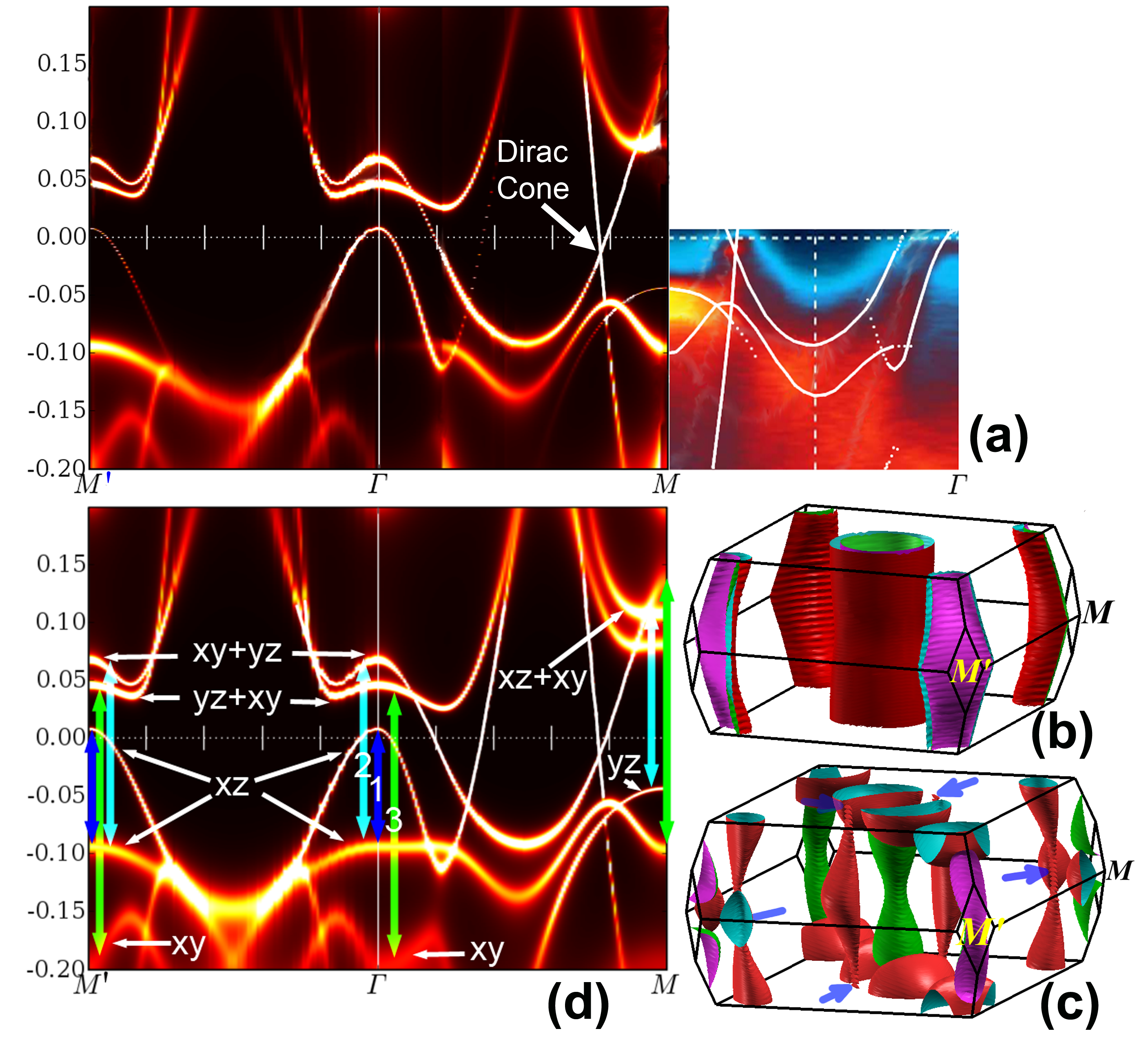}
  }
\caption{ARPES and Fermi surface of BaFe$_2$As$_2$.
(a) $A(\vk,\omega)$ in the $\Gamma$ plane in the path $M'\rightarrow \Gamma\rightarrow M\rightarrow \Gamma$ of the SDW state.
(See the locations of $M$ and $M'$ points in panel (b)).
In the path $M\rightarrow\Gamma$ we overlayed ARPES data from Ref.[\onlinecite{Richard}].
(b) Fermi surface in the PM state.
(c) Fermi surface in the SDW state, plotted in the PM Brillouin zone.
Blue arrows mark the position of the Dirac cones.
Note that the reciprocal vectors in the order state are $(1/2,-1/2,0)$ in the direction of the ferromagnetic ordering,
and $(1/2,1/2,1/2)$ in the direction of the antiferromagnetic ordering,
hence points $\Gamma$ and $M'$ are equivalent, while points $\Gamma$ and $M$ are not.
(d) $A(\vk,\omega)$ in the SDW state with shadow bands plotted by equal intensity for clarity.
Arrows mark the three types of optical transitions which give
rise to the three peaks in the optical conductivity.
}
\label{Akw}
\end{figure}

In Fig.\ref{Akw}(a), we show LDA+DMFT momentum-resolved electronic
spectra $A(\vk,\omega)$ in the SDW phase. This quantity is probed by
angle-resolved photoemission spectroscopy (ARPES). Figs.\ref{Akw}(b) and (c)
compare the Fermi surface of the PM and SDW states, displayed
in the PM Brillouin zone.

In the PM state, the topology of the Fermi surface is very
similar to LDA predictions~\cite{Singh} with three cylinders centered
at $\Gamma$ point and two at $M$ point. In the SDW phase, the Fermi
surface of the LSDA calculation (not shown) and LDA+DMFT is very
different.
The magnetic order reconstructs the Fermi surface into smaller more
three dimensional pockets. Out of three cylinders centered at $\Gamma$
point, one remains in the SDW phase. This cylinder does not intersect
the $\Gamma$ plane within LSDA (not shown), but has been clearly
identified in experiment. The other cylinders reconstruct into more
three dimensional pockets.

In the SDW state, there are two inequivalent directions between
$\Gamma$ and $M$, here named $M$ and $M'$, pointing along the
antiferromagnetic and ferromagnetic direction of the Fe-Fe bonds,
respectively. 
(see Fig.~\ref{Akw}(b)).
Graphene-like Dirac points were recently identified by
ARPES along the antiferromagnetic direction~\cite{Richard}. 
Fig.\ref{Akw}(a) shows that a crossing of two
bands occurs very near the Fermi level between $\Gamma$ and $M$, at
$3/4$ of the way, marked by a white arrow. The crossing is below the
Fermi level, hence the pocket is electron like. In Fig.~\ref{Akw}(c)
we mark the same tiny pocket by blue arrows, to show that it has
indeed a shape of a Dirac cone. There are two such symmetry related
Dirac cones in the $\Gamma$ plane and two in the $Z$ plane. Notice
that these cones appear only in the path between $\Gamma$ to $M$
(antiferromagnetic direction)  
and not in the $\Gamma$ to $M'$.

In the most right-hand part of Fig.~\ref{Akw}(a), we overlay our
results with ARPES measurements of Richard {\it et al.}\cite{Richard},
to emphasize common features. The overall position of the bands is in
very good agreement without any need of shift of the Fermi level or renormalization of
the bandwidth, in contrast to common need for shifts and
renormalization when comparing DFT-derived bands with ARPES.

The LDA+DMFT Fermi surface also has good agreement with the ARPES
measurement in the $Z$ plane by Shimojima {\it et al.}~\cite{Shimojima}. 
In particular, the red electron pockets
centered at $Z$, which have a two fold symmetry and mostly $xz$
character, were identified in Ref.\onlinecite{Shimojima}.

In Fig.~\ref{Akw}(d) we replot the momentum resolved electronic
spectra $A(\vk,\omega)$ without the SDW coherence factors, to enhance
the shadow bands. The arrows in this graph connect the features in the
electronic structure with the peaks in the optical conductivity. We
mark tree types of vertical transitions corresponding to the three
peaks in Fig.~\ref{optical-conductivity}.
The first shoulder comes primarily from transitions within the $xz$
orbital, namely between the flat band around $-0.1\,$eV and the hole
pocket at $\Gamma$ and $M$. These transitions are between a shadow
band and a non-shadow band, hence they appear only in the SDW phase.
The second peak in optics, marked by cyan arrow, comes primarily from
transitions between the $xz$ and $xy$ orbitals, with some transitions
between non-shadow bands only, visible also in the PM state, and some
additional transitions between a shadow and non-shadow band. Finally,
the third peak comes mostly from transitions between the $xy$ and $yz$
orbitals, and mostly from transitions between a shadow to non-shadow
band.

In correlated materials new physics, such as superconductivity, spin
and orbital polarization, emerge from the competition between Coulomb
interaction and kinetic energy.

A unique physical characteristics of iron arsenic materials is that
the kinetic energy loss in the SDW phase is compensated by a gain in
Hund's rule coupling energy.
Comparison of the SDW and PM histograms, describing the probability of
different iron configuration in the solid, shows that in the SDW state
the high spin states become more probable. This results in an overall
gain of the Hund's rule coupling energy of about 500$\,$K per Fe.
Overall kinetic energy is lost in the SDW state for about $300\,$K per
Fe, resulting in a net energy gain of about $200\,$K per Fe.
This is different from classical SDW transition where kinetic energy
is compensated by the reduction of the Hubbard correlations.

The competition between kinetic and correlation energy takes different
forms at different energy scales, and results in an unusual energy
dependence of the spin and orbital polarization.
Spin polarization affects most strongly the electrons far below the
Fermi level. For example, the exchange splitting - as determined from
the frequency dependent potential (self-energy) at high frequency - is
three times larger than at zero frequency. This high frequency
regime, is governed by the strong Hund's rule coupling on iron atom,
enhancing magnetic moment.

At low energies, in the SDW state, well defined quasiparticles form,
and the residual Hund's rule coupling between these quasiparticles is
weak. To minimize the kinetic energy loss in the SDW phase, the
quasiparticles propagate mainly along the antiferromagnetic $x$
direction, the direction which is not blocked by the Pauli exclusion
principle. This generates strong orbital polarization, but only at low
energy, where the quasiparticles are well formed, and the effective
Hund's rule coupling is weakest. On the contrary, the overall orbital
polarization is weak.

Our finding that spin polarization is larger at high energy while the
orbital polarization is most pronounced at low energies leads to
definite predictions for the anisotropy of the optical conductivity
and can be tested also by STM.

Energy dependent polarizations and the enhancement of coherence of the
low energy quasiparticles in the SDW phase, can only be described by
frequency dependent potentials and Weiss fields, as is done in
DMFT. This explains the failure of static mean
field theories such as LDA to capture both the correct moment, which
lives at high energies, and the low energy spectra.
In a renormalization group picture of this material, one observes a
different strength of the Hund's rule coupling at different energy
scales. At high energy, Hund's rule coupling is very strong, while it
fades away at low energy, but gives an imprint on the massive and
anisotropic low energy quasiparticles.
This is central for a proper description of the magnetic phase, and is
likely to be important for the mechanism of the unconventional
superconductivity in these materials.

\textbf{METHOD}\\
To show that the origin of the anisotropy is electronic rather than
structural, we use the experimental lattice constants and internal
coordinates of the paramagnetic phase~\cite{Ba-lattice}. We use the
continuous time quantum Monte Carlo as the impurity solver and charge
self-consistent version of LDA+DMFT, described in detail in
Ref.\onlinecite{Haule-DMFT}. We use \textit{ab initio} determined Coulomb
interaction strength $U=5.0\,$eV and $J=0.7\,$eV~\cite{Kutepov}, and
temperature $T=72.5\,$K.
To explore the sensitivity of the magnetic moment to the strength of
the Coulomb interaction, we performed calculations for other values of
$U$ and $J$ around the \textit{ab initio} values.  We found that the size of
the magnetic moment can be well parameterized by the simple formula
$m=(0.4\, U + 7.2\, J-6.1eV)\mu_B/$eV.  Hence, magnetization is most sensitive to
the value of the Hund's coupling $J$, rather than $U$.

\textbf{Supplementary Information} is linked to the online version of the paper at www.nature.com/nature.

\textbf{Acknowledgments} ZPY is grateful to Chuck-Hou Yee for help at the initial stage of this
project. We are grateful to D. Basov for fruitful discussions. ZPY and GK were supported by NSF DMR-0906943, KH was supported
by NSF DMR-0746395 and Alfred P. Sloan foundation.

\textbf{Author Information:} Correspondence and requests for materials should be addressed to ZPY at yinzping@physics.rutgers.edu.

\newpage
\mbox{}

\newpage

{\bf Magnetism and Charge Dynamics in Iron Pnictides\\
Supplementary Information
}

Z. P. Yin, K. Haule and G. Kotliar \\
Department of Physics and Astronomy, Rutgers University, Piscataway, NJ 08854, United States.

\vskip 5mm
The two types of polarization, spin and orbital, are dominant at
different energy scales.  Spin polarization is stronger at higher
energy (far below the Fermi level) while orbital polarization is
strong at low energy (close to the Fermi level).

\begin{figure}[!ht]
\centering{
  \includegraphics[width=0.95\linewidth]{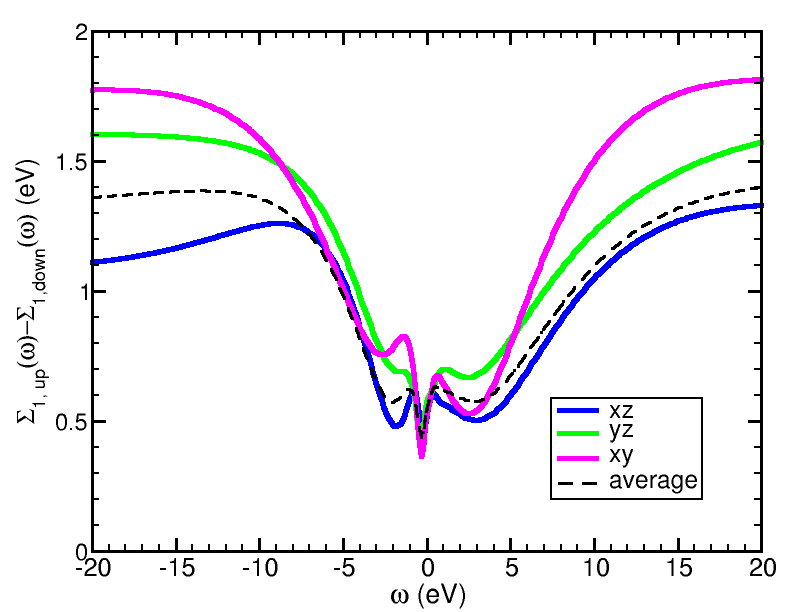}
  }
\caption{Spin polarization of the frequency dependent self energy (real part)
$\Sigma_{1, \uparrow}(\omega)-\Sigma_{1, \downarrow}(\omega)$
of the Fe $3d_{xz}$, $3d_{yz}$, $3d_{xy}$ orbitals,
and the average over all five $3d$ orbitals in the SDW state of BaFe$_2$As$_2$ calculated by LDA+DMFT.
}
\label{spin-polarization}
\end{figure}

To demonstrate that the spin polarization is stronger at high energy,
we plot in Fig.~\ref{spin-polarization} the exchange splitting of
iron $3d$ orbitals, defined by
$$\Sigma_{1, \uparrow}(\omega)-\Sigma_{1, \downarrow}(\omega).$$ Here
$\Sigma(\omega)$ is the frequency dependent potential (self-energy), which is added
to the one particle Hamiltonian, to produce the many body spectra.
The high frequency limit of the exchange splitting is on average three
times of its zero frequency value. Hence, the magnetic moment comes
mainly from the high energy region, where the exchange splitting is
large.

At energies within the SDW gap, the kinetic energy is dominant. To
minimize the kinetic energy in the magnetically ordered state,
electrons create a highway in the direction in which spins are
antiferromagnetically ordered, while they remain slow in the
ferromagnetic direction due to Pauli blocking. This gives rise to
strong orbital polarization of the iron $3d$ orbitals at energies
close to the Fermi level.

\begin{figure}[!ht]
\centering{
  \includegraphics[width=0.95\linewidth]{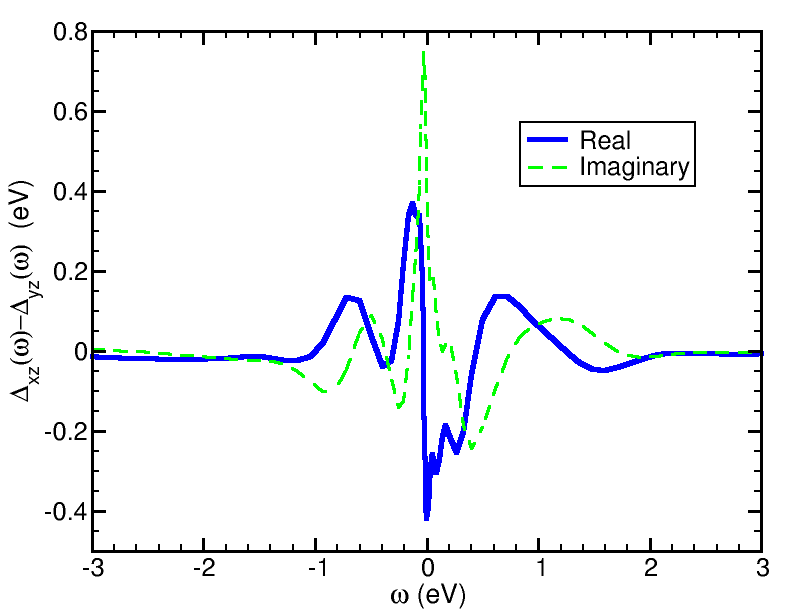}
  }
\caption{Orbital polarization of the frequency dependent hybridization $\Delta_{xz}(\omega)-\Delta_{yz}(\omega)$
of the Fe $3d_{xz}$ and $3d_{yz}$ orbitals summing over both spin channels in the SDW state of BaFe$_2$As$_2$ calculated by LDA+DMFT.
}
\label{orbital-polarization}
\end{figure}

The orbital polarization is related to the frequency dependent Weiss
field $\Delta(\omega)$, which describes the hybridization of the iron
atom with the rest of the system. In Fig.~\ref{orbital-polarization}
we plot the difference of the hybridization between the $xz$ and $yz$
orbital, i.e., $\Delta_{xz}(\omega)-\Delta_{yz}(\omega)$.  In the
paramagnetic state, the $xz$ and $yz$ hybridization are equal due to the
tetragonal crystal structure, but they become different in the SDW
state. As shown in Fig.~\ref{orbital-polarization} the orbital
polarization acquires finite value only in the region of dominantly
coherent spectra (within $1.5\,$eV from the Fermi level), and gets
strongly enhanced within the SDW gap, where the quasiparticles are
well defined.

\end{document}